\documentclass[twocolumn]{aa}
\usepackage{graphics}
\usepackage{txfonts}

\begin{document}
\title{Observational determination of the time delays in gravitational lens
system Q2237+0305}

\author{V.Vakulik \inst{1}
       \and  R.Schild \inst{2}
       \and  V.Dudinov \inst{1}
       \and  S.Nuritdinov \inst{3} 
       \and  V.Tsvetkova \inst{4}
       \and  O.Burkhonov \inst{3}
       \and  T.Akhunov \inst{3}}     
\offprints{V.Vakulik, \email{vakulik@astron.kharkov.ua}}

\institute{Institute of Astronomy of Kharkov National University, Sumskaya
           35, 61022 Kharkov, Ukraine         
\and  Center for Astrophysics, 60 Garden Street, Cambridge, MA
           02138, U.S.A.          
\and  Ulugh Beg Astronomical Institute of Ac.Sci. of Uzbekistan,
           Astronomicheskaya 33, 700052, Tashkent, Republic of 
           Uzbekistan   
\and  Institute of Radio Astronomy of Nat.Ac.Sci. of Ukraine, 
           Krasnoznamennaya 4, 61002, Kharkov, Ukraine}
           
\date {Received... / Accepted...}  
\titlerunning{Time delays in Q2237}

\abstract{We present new brightness monitoring observations of the 
4 components of gravitationally lensed system Q2237+0305, 
which show detection of an intrinsic quasar brightness fluctuation 
at a time of subdued microlensing activity, between June 27 and 
October 12, 2003. These data were used to determine the time delays 
between the arrivals of the four images. The measured delays are 
$\tau_{BA}\approx-6$, $\tau_{CA}\approx35$, and $\tau_{DA}\approx2$ 
hours, so they confirm that the long history of brightness monitoring 
has produced significant detection of microlensing. However the error 
bars associated with the delays, of order 2 days, are too large to 
discriminate between competing macro-imaging models. Moreover, our 
simulations show that for the amplitude of this intrinsic fluctuation 
and for photometric errors intrinsic to optical monitoring from our 
1.5-m telescope or from the OGLE monitoring, a daily sampled 
brightness record cannot produce reliable lags for model discrimination. 
We use our simulations to devise a strategy for future delay 
determination with optical data. Nevertheless, we regard these first 
estimates to be significant, since they are the first direct 
measurements of time delays made for this system from ground-based 
observations in the visual wavelengths. Our results provide the most 
convincing confirmation of the gravitational-lens nature of Q2237+0305, 
and give observational justification to the extensive literature which 
attributes the quasar's previously observed brightness fluctuations to 
microlensing.

}

\maketitle

\section{Introduction}

The quadruple Q2237+0305 gravitationally lensed quasar 
system was among the first discovered (Huchra et al. \cite{huch85}, 
Yee \cite{yee88}), and has frequently been cited as an important system 
in which microlensing by stars in the lens galaxy should be readily 
observed with events having duration of a few hundred days. However 
microlensing studies have ordinarily required determination of time 
delays to permit discrimination between intrinsic quasar fluctuations 
and microlensing-induced brightness fluctuations. In the case of 
Q2237+0305, models of the macro-imaging from the earliest times 
(e.g. Schneider et al. \cite{schn88}, Rix et al. \cite{rix92}, 
Wambsganss \& Paczynski \cite{w-p94}), as well as the most recent 
and most popular model by Schmidt, Webster \& Lewis (\cite{swl98}) 
all suggested that the time delays should be short, of order a few 
hours up to 1 day, but this prediction has never been confirmed by 
observations. This is because of the high optical density to 
microlensing for all the four quasar components that their brightness 
changes are unceasing and significantly uncorrelated. Till recently, 
the available light curves of the Q2237+0305 components did not reveal 
any region of correlated brightness variations, which could be 
attributed to the quasar intrinsic brightness fluctuation, with one 
exception; \O stensen et al.(\cite{ost96}) noted a synchronous 
brightening of all four components in 1994, - in the very end of their 
monitoring program with the Nordic Optical Telescope. 
It was interpreted as a possible intrinsic variation of the source, 
reaching 0.0006 magnitudes per day. The authors found this to be 
insufficient for determining the time delays, but noted a rather high 
correlation between the components. A single measurement of time delay 
for Q2237+0305 exists so far, made by Dai et al. (\cite{dai03}) from 
the Chandra X-ray observations. They reported a 2.7 hours time delay 
between images A and B, with image A leading. 
 
When the first Q2237 microlensing fluctuations were announced in 1989 by 
Irwin et al. (\cite{irw89}), the Q0957 time delay had already been 
measured by Schild \& Cholfin (\cite{s-ch86}) and confirmed by Vanderriest 
et al. (\cite{van89}). Both authors reported evidence of microlensing in 
the Q0957 system (Grieger, Kayser, and Refsdal \cite{gri86}; Vanderriest 
et al. \cite{irw89}), but the claim was not completely accepted because 
the time delay was challenged by Lehar et al. (\cite{leh92}) from radio 
data and by Press et al. (\cite{pre92}) from re-analysis of the optical 
data. However redetermination of the Q0957 time delay by Schild \& Thomson 
(\cite{s-th97}), Pelt et al. (\cite{pelt96}) and Kundic et al. 
(\cite{kun97}) firmly established the original Schild \& Cholfin 
(\cite{s-ch86}) time delay and, thus also, the first discovered 
microlensing event, evidently caused by a star in the lens galaxy 
(Schild \& Smith 1991).

During the same time period, numerous events were detected in the Q2237 
system and discussed in the context of microlensing even though the time 
delay had not been measured, but was presumed to be short, (Corrigan 
\cite{cor91}; Wambsganss, Paczynski, and Schneider \cite{wps90}). 
Thus it would be important to observationally confirm this prediction 
in order to validate the microlensing inferences. However because the 
microlensing of the individual images is seen nearly continuously, it 
has been difficult to date to detect an intrinsic quasar fluctuation 
from which the time delays could be determined. A similar conundrum 
has been discussed in the case of Q0957 by Colley and Schild 
(\cite{c-s03}); because of the microlensing it has been difficult to 
determine an accurate time delay, but without an accurate time delay 
it has been difficult to study the microlensing.

We report below the observation of an unambiguous Q2237 intrinsic 
quasar fluctuation seen in all four images. In section 2 we report 
our observations with the 1.5-m Maidanak telescope, and their 
reduction to form the four time series records. In section 3 we 
determine the delays between the arrivals of the four images by 
two methods, and in the final sections we discuss our results 
in the context of intrinsic quasar fluctuations observed in other 
quasars, and in the context of the microlensing and macro-lensing 
models.

\section{Observations and data reduction}

CCD images of the Q2237+0305 gravitationally lensed quasar were taken 
with the Maidanak 1.5-m telescope in the framework of an international 
cooperative program of monitoring gravitational lenses, launched
in 1997, (Dudinov et al. \cite{dud00}, Vakulik et al. \cite{vak04}).
An LN-cooled BroCam CCD camera with a SITe ST 005A chip was available 
in the f/8 focal plane, giving a scale of 0.26\arcsec/pix. Magnitude 
measurements of the four quasar components with the corresponding 
uncertainties are shown in Table~\ref{2003_phot}, where the Julian 
dates and the seeing conditions (FWHM values) are also indicated.

Because of some problems with the telescope tracking system, we used 
rather short exposures, which do not exceed 3 min. Therefore, to provide 
the highest photometric accuracy, the Q2237 images were 
usually taken in series, consisting of 4 to 12 frames. The frames were 
averaged before being subjected to photometric processing, thus ensuring 
the accuracy of about $0.008^m$ for A, $0.018^m$ for B, $0.016^m$ for C, 
and $0.020^m$ for the D component. These are the mean errors, averaged 
over the entire dataset, while the error values presented in 
Table~\ref{2003_phot} for every date were obtained from a comparison of 
the photometry of individual frames in the series, and may be regarded 
as an adequate estimate of the random error inherent in a particular 
quasar image component series. 

\begin{table*}
\renewcommand{\arraystretch}{0.7}
\caption{Photometry of Q2237+0305 A,B,C,D in the $R$ band from 
observations with the Maidanak 1.5-m telescope in June-October 2003.} 
\label{2003_phot}
\begin{center}
\begin{tabular}{cccccccc}
\hline
 \noalign{\smallskip}
     &    &               &            &           &         &    & Number \\
Date & JD &   A           &    B       &     C     &      D  &FWHM    & of\\
     &(2452000+)&         &            &           &         &(arcsec)&frames\\
\hline
\hline     
jun27&818&17.055$\pm$0.003&18.549$\pm$0.029&18.285$\pm$0.013&18.392$\pm$0.031&1.0&4\\
jun28&819&17.071$\pm$0.014&18.615$\pm$0.030&18.282$\pm$0.012&18.389$\pm$0.014&1.0&4\\
jun29&820&17.069$\pm$0.013&18.593$\pm$0.026&18.286$\pm$0.015&18.450$\pm$0.021&1.0&4\\
jul01&822&17.076$\pm$0.005&18.604$\pm$0.023&18.272$\pm$0.021&18.437$\pm$0.042&0.9&7\\
jul04&825&17.049$\mp$0.008&18.533$\mp$0.032&18.287$\mp$0.025&18.413$\mp$0.042&1.0&4\\
jul07&828&17.053$\mp$0.010&18.574$\mp$0.004&18.260$\mp$0.005&18.415$\mp$0.008&0.9&4\\
jul08&829&17.070$\pm$0.009&18.590$\pm$0.054&18.274$\pm$0.032&18.424$\pm$0.109&1.0&4\\
jul10&831&17.052$\pm$0.014&18.598$\pm$0.035&18.310$\pm$0.034&18.414$\pm$0.036&1.2&6\\
jul14&835&17.077$\pm$0.005&18.605$\pm$0.025&18.305$\pm$0.023&18.422$\pm$0.045&1.1&8\\
jul19&840&17.086$\pm$0.006&18.595$\pm$0.029&18.280$\pm$0.023&18.436$\pm$0.021&0.9&6\\
jul26&847&17.077$\pm$0.006&18.594$\pm$0.011&18.289$\pm$0.020&18.439$\pm$0.008&1.0&4\\
jul28&849&17.057$\pm$0.018&18.622$\pm$0.015&18.280$\pm$0.057&18.446$\pm$0.049&1.2&4\\
jul30&851&17.056$\pm$0.003&18.585$\pm$0.005&18.269$\pm$0.022&18.430$\pm$0.034&1.1&4\\
jul31&852&17.069$\pm$0.001&18.566$\pm$0.012&18.284$\pm$0.015&18.413$\pm$0.012&0.9&6\\
aug05&857&17.073$\pm$0.007&18.564$\pm$0.016&18.262$\pm$0.012&18.416$\pm$0.016&1.2&12\\
aug06&858&17.070$\pm$0.006&18.592$\pm$0.009&18.246$\pm$0.014&18.425$\pm$0.012&0.9&4\\
aug08&860&17.070$\pm$0.004&18.570$\pm$0.007&18.259$\pm$0.008&18.413$\pm$0.012&0.9&9\\
aug09&861&17.072$\pm$0.005&18.578$\pm$0.013&18.270$\pm$0.016&18.437$\pm$0.014&0.8&6\\
aug10&862&17.069$\pm$0.005&18.577$\pm$0.017&18.276$\pm$0.015&18.435$\pm$0.020&0.9&4\\
aug11&863&17.066$\pm$0.018&18.606$\pm$0.046&18.236$\pm$0.036&18.440$\pm$0.043&1.0&5\\
aug14&866&16.925$\pm$0.009&18.454$\pm$0.059&18.201$\pm$0.022&18.437$\pm$0.039&0.9&4\\
aug15&867&17.019$\pm$0.006&18.534$\pm$0.026&18.235$\pm$0.030&18.362$\pm$0.019&1.1&4\\
aug16&868&17.089$\pm$0.006&18.622$\pm$0.018&18.288$\pm$0.012&18.427$\pm$0.022&1.1&4\\
aug21&873&17.062$\pm$0.010&18.543$\pm$0.015&18.242$\pm$0.017&18.385$\pm$0.018&0.8&4\\
aug22&874&17.032$\pm$0.004&18.555$\pm$0.012&18.245$\pm$0.007&18.399$\pm$0.015&0.9&4\\
aug23&875&17.048$\pm$0.005&18.567$\pm$0.005&18.249$\pm$0.012&18.400$\pm$0.019&1.0&6\\
aug24&876&17.038$\pm$0.009&18.533$\pm$0.015&18.246$\pm$0.006&18.413$\pm$0.014&1.0&4\\
aug25&877&17.037$\pm$0.005&18.546$\pm$0.012&18.237$\pm$0.012&18.401$\pm$0.022&1.0&4\\ aug26&878&17.017$\pm$0.006&18.525$\pm$0.006&18.222$\pm$0.007&18.380$\pm$0.017&0.8&4\\
aug27&879&17.039$\pm$0.007&18.555$\pm$0.023&18.258$\pm$0.010&18.388$\pm$0.011&0.8&4\\
aug28&880&17.026$\pm$0.005&18.541$\pm$0.019&18.222$\pm$0.004&18.394$\pm$0.011&1.0&6\\
sep01&884&17.006$\pm$0.010&18.535$\pm$0.022&18.193$\pm$0.017&18.390$\pm$0.009&1.0&4\\
sep02&885&16.978$\pm$0.032&18.583$\pm$0.026&18.183$\pm$0.059&18.373$\pm$0.047&1.5&4\\
sep05&888&16.981$\pm$0.012&18.515$\pm$0.029&18.219$\pm$0.024&18.366$\pm$0.036&1.1&4\\
sep06&889&16.996$\pm$0.016&18.491$\pm$0.065&18.169$\pm$0.057&18.282$\pm$0.019&1.1&4\\
sep12&895&16.826$\pm$0.015&18.567$\pm$0.227&18.089$\pm$0.023&18.394$\pm$0.153&1.4&4\\
sep13&896&16.991$\pm$0.003&18.533$\pm$0.031&18.215$\pm$0.008&18.327$\pm$0.033&0.9&3\\
sep15&898&16.924$\pm$0.024&18.450$\pm$0.039&18.103$\pm$0.046&18.340$\pm$0.051&1.5&8\\
sep16&899&16.923$\pm$0.006&18.513$\pm$0.028&18.154$\pm$0.033&18.395$\pm$0.048&1.2&4\\
sep17&900&16.975$\pm$0.003&18.471$\pm$0.007&18.201$\pm$0.004&18.310$\pm$0.014&0.9&12\\
sep18&901&16.953$\pm$0.008&18.499$\pm$0.010&18.157$\pm$0.011&18.358$\pm$0.014&0.9&8\\
sep19&902&16.946$\pm$0.004&18.473$\pm$0.010&18.142$\pm$0.007&18.319$\pm$0.012&1.1&8\\
sep20&903&16.937$\pm$0.003&18.466$\pm$0.004&18.151$\pm$0.012&18.305$\pm$0.013&0.8&4\\
sep21&904&16.923$\pm$0.005&18.465$\pm$0.010&18.123$\pm$0.016&18.308$\pm$0.011&1.2&8\\
sep24&907&16.924$\pm$0.006&18.487$\pm$0.011&18.144$\pm$0.003&18.325$\pm$0.011&1.0&6\\
sep25&908&16.933$\pm$0.004&18.470$\pm$0.013&18.135$\pm$0.010&18.331$\pm$0.010&0.9&8\\
sep27&910&16.928$\pm$0.008&18.503$\pm$0.019&18.139$\pm$0.025&18.311$\pm$0.016&1.2&6\\
sep29&912&16.917$\pm$0.007&18.440$\pm$0.010&18.115$\pm$0.010&18.329$\pm$0.010&1.2&7\\
sep30&913&16.927$\pm$0.002&18.469$\pm$0.003&18.133$\pm$0.007&18.313$\pm$0.004&0.8&8\\
oct01&914&16.932$\pm$0.005&18.470$\pm$0.011&18.138$\pm$0.015&18.323$\pm$0.006&0.9&4\\
oct02&915&16.929$\pm$0.003&18.476$\pm$0.008&18.126$\pm$0.006&18.329$\pm$0.004&0.8&8\\
oct11&924&16.938$\pm$0.003&18.479$\pm$0.011&18.137$\pm$0.009&18.315$\pm$0.009&0.9&8\\
oct12&925&16.920$\pm$0.005&18.433$\pm$0.009&18.117$\pm$0.007&18.287$\pm$0.012&1.0&4\\
 
\noalign{\smallskip} \hline
\end{tabular}
\end{center}
\end{table*}

The algorithm of photometric image processing is described in great 
details in (Vakulik et al., \cite{vak04}), where the absence of 
seeing-dependent systematic photometric errors for images with PSF 
values up to $1.4\arcsec$ is claimed. It is interesting to note 
as well, that the errors indicated in Table~\ref{2003_phot} reveal 
no seeing dependence for PSFs up to $1.2\arcsec$. However, one can 
easily notice much larger errors for three dates, when the seeing 
was $1.4\arcsec$ to $1.5\arcsec.$
          
\begin{figure}
\resizebox{\hsize}{8cm}{\includegraphics{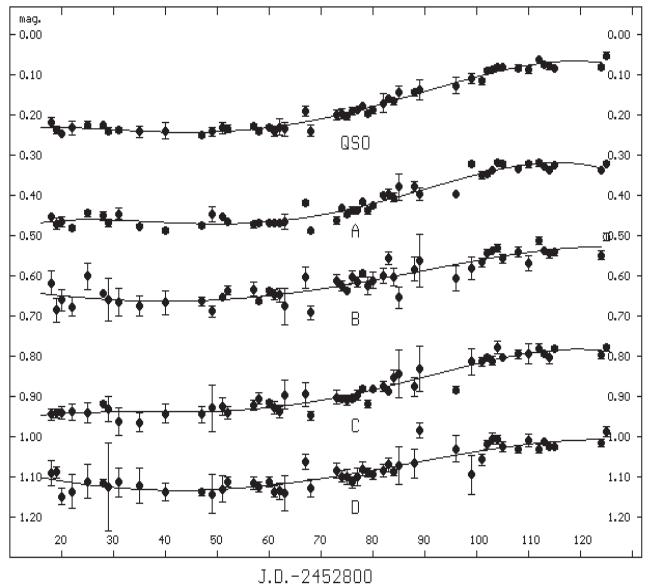}}
 \caption{Synchronous variations of brightnesses of the A-D images 
 observed in Q2237+0305 between June 27 and October 12, 2003 with the 
 Maidanak 1.5-m  telescope. R magnitudes shifted arbitrarily along the 
 vertical axis are  shown. The upper curve is an integral light curve
 of the four quasar images (see the text for more details).}
 \label{l_curves}
\end{figure}

The Q2237+0305 A,B,C,D light curves are shown in Fig.\ref{l_curves}. 
To better illustrate their similarity, they are arbitrarily shifted 
along the magnitude axis. The upper curve is a total light curve 
obtained from  the aperture photometry of the central part of the 
object within a diaphragm of 3\arcsec  radius, and with the subsequent 
subtraction of the intensity contribution from the galaxy. 
Polynomial fits (fifth order polynomials were used) are also shown. 
All light curves are well correlated, thus demonstrating extremely 
low microlensing activity during this time period, while the quasar 
revealed rather high variability, - $0.15^m$ between the Julian 
dates 860 and 910, i.e. about $0.003^m$ per day. This is almost five 
times more rapid brightness change as compared to the measurements by 
\O stensen et al. (\cite{ost96}) in 1994. Similar synchronous 
brightness variations can be seen in the $V$ light curves obtained by 
the OGLE group for the same time period, 
(http://www.astrouw.edu.pl~ogle/ogle3/huchra.html),
though the authors do not note similarity between the light curves of 
the components. A quantitative measure of similarity between the light 
curves can be seen from Table~\ref{corr_ind}\ref{l_curves}, where the 
corresponding correlation indices are shown for our photometry and for 
the OGLE data. Since the expected time delays are very short, the 
table justifies an attempt to undertake calculation of the time delays, 
which is possible thanks to subdued, though non-zero microlensing 
activity. 

\begin{table}
\caption{Correlation indices between the light curves of the Q2237
components calculated for our data, and for the OGLE photometry. 
Quantities in the last line and column of the first table, both 
marked as QSO, are correlation indices between a particular light 
curve and the integral light curve of the quasar.} 
\label{corr_ind}
\begin{center}
\begin{tabular}[h]{cccccc}
\raisebox{1ex}{\it{Our data}}& & & & &\\
\hline
 \noalign{\smallskip}
 Component & A & B & C & D & QSO\\
 \hline
A  &   1   & 0.899 & 0.960 & 0.874 & 0.978 \\
B  & 0.898 &   1   & 0.888 & 0.887 & 0.924 \\
C  & 0.960 & 0.888 &   1   & 0.874 & 0.968 \\
D  & 0.874 & 0.887 & 0.874 &   1   & 0.922 \\
QSO& 0.978 & 0.924 & 0.968 & 0.922 &  1    \\
\hline
&&&&&\\
\raisebox{1ex}{\it {OGLE data}}& & & & &\\
\hline
 \noalign{\smallskip}
Component& A & B   & C     & D     &\\
 \hline
A  &   1   & 0.945 & 0.953 & 0.798 &  \\
B  & 0.945 &   1   & 0.914 & 0.754 &  \\
C  & 0.953 & 0.914 &   1   & 0.726 &  \\
D  & 0.798 & 0.754 & 0.726 &   1   &  \\
\hline
\end{tabular}
\end{center}
\end{table}

\begin{figure}
\resizebox{\hsize}{7cm}{\includegraphics{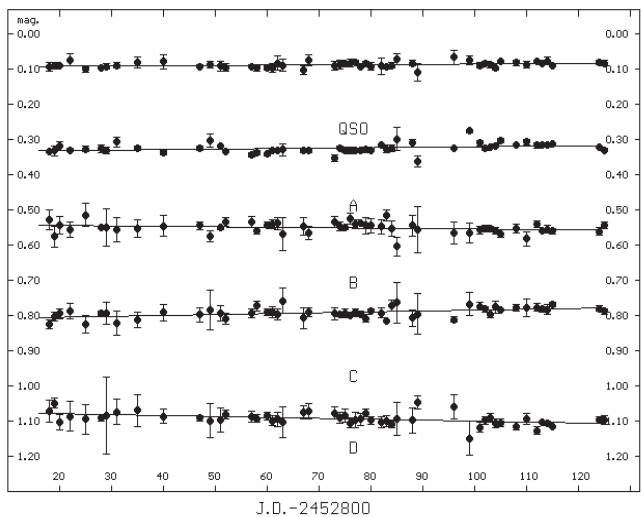}}
 \caption{Residual variations of brightnesses of the A-D images  
 taken from the light curves of June 27 - October 12, 2003 (Fig.1) 
 by subtracting the integral quasar light curve. These residuals may
 be treated as a result of microlensing.}
 \label{mic_trends}
\end{figure}

It is interesting to estimate the level of possible microlensing 
activity during this time period. In Fig.\ref{mic_trends}, 
variations of magnitudes of individual quasar components with 
respect to their integral light curve are shown, which are probably 
caused by microlensing. The figure illustrates, that the microlensing 
variations are rather slow and may be regarded as linear trends 
during the 100-day monitoring period. The mean slopes fitted to 
these microlensing residuals were -0.000136, 0.000125, -0.000251, 
and 0.000264 magnitudes per day for the A, B, C and D components, 
respectively. These are important quantities, which we will need 
in the next section to analyse uncertainties of the time delays 
estimates.

We tried to fit the observed brightness changes by polynomials of 
the 3-rd, 4-th, 5-th, and 6-th orders, and found that there is no 
justification to use the polynomial degree higher than four in 
calculations, since scatter of observational data points from the 
fitting curves decreases very slowly as the polynomial order grows. 
For example, the fourth-order polynomial gives the RMS deviations for 
the datapoints of $0.0148^m$, $0.0228^m$, $0.0161^m$, $0.0206^m$, 
and $0.0098^m$ for the A, B, C and D components and the quasar, 
respectively, whereas with the fifth-order polynomial, the RMS 
deviations are $0.0147^m$, $0.0227^m$, $0.0159^m$, $0.0204^m$, 
and $0.0098^m$. 

This quantities may be regarded as an upper estimate for the 
actual errors of our photometry, while the values $0.008^m$, 
$0.018^m$, $0.016^m$, and $0.020^m$ indicated in the second paragraph 
of this section, should be regarded as a measure of the intrinsic 
convergence of the photometry for images A, B, C and D, respectively. 
Thus the data presented in 
Table~\ref{2003_phot} and shown in Fig.\ref{l_curves} are accurate 
enough to attempt calculations of time delays. Table~\ref{corr_ind} 
contains an additional argument in favour of such an attempt, - the 
light curves of all the four quasar images are well correlated. 
 
Nevertheless, to be sure that such featureless light curves are 
really capable of providing reliable estimates of time delays, 
to investigate the effects of photometry errors and residual 
microlensing effects, and to obtain consistent estimates of the 
time delay uncertainties, numerical simulations involving over 
2000 statistical trials were undertaken. The simulation results 
will be briefly described at the end of the next section, after 
the methods used to determine time delays are presented.
 
\section{Time Delay Estimation}

The detection of correlated variations of brightnesses of the 
Q2237 components is important in itself, since it is the most 
convincing evidence of the gravitational lensing nature of the 
system. The deeper value of these observations is in the 
possibility to estimate the time delays between the brightness 
fluctuations seen in different lensed quasar images.

Estimation of time delays from the actual light curves, with their
inevitable uneven sampling and poor statistics, is a rather difficult 
problem. A great variety of approaches has been elaborated, 
with the most important ones being analysed by Gil-Merino 
et al. (\cite{gil02}). All of these algorithms are devised for the 
situation when the characteristic timescale of the quasar 
brightness variations and sampling time intervals of observations 
are much less than the expected time delays. The existing 
macrolensing models for this system predict time delays which 
do not exceed a day. Therefore, we have a somewhat atypical situation, 
- we are trying to estimate the time delays from brightness 
variations with the characteristic timescale much larger than the 
expected value of the time delay. Moreover, we used photometry 
data averaged within a night, and thus the sampling time interval 
exceeds the expected values of time delays too.

Taking all this into account, we used the following approach. We 
admit that the quasar brightness variations are rather slow and 
can be represented accurately enough by a fourth or fifth order 
polynomial. Then, representing brightness variations of a component, 
e.g. A, with a smooth curve $f_A(t)$, we may calculate a normalized 
cross-correlation function (NCCF) of this smooth curve with the 
actual light curve of another component, say, B, represented by 
discrete data points:
\begin{equation}
 \label{cr_cor}
 k(\tau)={{{1\over n}{\sum^n_{i=1}[f_A(t_i+\tau)-\overline{f}_A(\tau)]
 [m_B(t_i)- \overline{m}_B]}}\over {(\sigma^2_f\sigma^2_m})^{1/2}},
\end{equation}
where $m_B(t_i)$ are discrete brightness samples at the time moments 
$t_i$, $f_A(t_i+\tau)$ are corresponding values of the approximating 
polynomial at the time moments $t_i+\tau$, $\overline{f}_A$ and 
$\overline{m}_B$ are corresponding average values, and $\tau$ is the 
time shift. The average values $\overline{f}(\tau)$, $\overline{m}$ 
and variances $\sigma^2_f$,$\sigma^2_m$ are calculated according to 
formulas:
\begin{equation}
\label{average}
\overline{f}(\tau)={1\over n}\sum_{i=1}^n f(t_i+\tau), \quad 
\overline{m}={1\over n}\sum_{i=1}^n m(t_i);
\end{equation}

\begin{equation}
 \label{variance}
 \sigma^2_f(\tau)=\sum^n_{i=1}{{[f(t_i+\tau)-
 \overline{f}(\tau)]^2}\over{n-1}}, \quad
 \sigma^2_m=\sum^n_{i=1}{{[m(t_i)-\overline{m}]^2}\over{n-1}}.
\end{equation}

With such an approach, we expect to find cross-correlation peaks at 
the cosmological lags. We adopted this procedure because it is less 
sensitive to bias by any false peaks for 0 lag that can result from 
the kinds of correlated brightness fluctuations at 0 lag when 
working with discrete data available from optical monitoring. These 
correlated photometry errors are expected from the inevitable 
residual sensitivity of the photometry to seeing effects, that  
is often observed, (e.g. Vakulik et al. \cite{vak04}). The effect 
of such "zero lag" is also discussed in Colley et al.(\cite{c-sch03}) 
and interpreted as a "frame-to-frame correlation error in the 
photometry".

The NCCF, calculated in this way, is a smooth function, with one 
maximum, which changes very slowly, and does not have other local 
extrema at least within $-10<\tau<10$ days. It allows us to 
unambiguously interpret the location of the maximum of this function 
as an optimal value of the lag, which we accept as the time delay 
between the brightness variations of two components. In this way we 
can estimate the time delays for the light curves of all the 
components with respect to the fitting curve of the A component.

Similar calculations can be made with respect to the fitting 
curves of other components, and thus a complete set of the time delay 
estimates will be obtained. Such estimates cannot be regarded as 
independent ones however, since they are affected with the same 
random errors of light curves of the "template" components, while 
their differences are caused by the difference between the 
approximating curves used.

We also propose a more general approach, consisting in a joint 
fitting of all the four light curves by a single polynomial, with 
a normalizing factor, magnitude shift, and time shift of each 
particular light curve being treated as unknown parameters and found 
from a condition:
\begin{equation}
 \label{joint}
 \sum_j\sum_i[k_j f(\vec a,t_i,\tau_j)+a_{0,j}-m_j(t_i)]^2=min,
\end{equation}
\begin{equation}
f(\vec a,t_i,\tau_j)=\sum_{n=1}^N a_n(t_i+\tau_j)^n.
\end{equation}

Here, $i$ is a number of a data point in the light curve of the $j$-th
component, $k_j$ is a normalizing factor for the $j$-th light curve, 
and $\tau_j$ is the unknown time delay for the $j$-th component. 
The function $f(\vec a,t_i,\tau_j)$ is an $N$-th order polynomial 
without the zeroth order term which is represented for each component 
separately as the magnitude shift $a_{0,j}$. 

These two approaches, - which we marked as NCCF and JFM (joint fitting 
method), -  have been applied to the photometry data of June-October 
2003, presented in Table 5 and shown in Fig. \ref{l_curves}. The 
estimates of time delays of the B, C and D images with respect to 
image A are presented in Table~\ref {time_delays}. Column 2 contains 
the estimates, taken with the NCCF method. They are the results of 
averaging the time delay estimates calculated  with respect to the 
fitting curves of all the components separately. In column 3, results 
of applying the JF method are shown. A comparison of these two columns  
illustrates the validity of both methods, - they produce almost the 
same time delay values. 

If, in addition to the quasar intrinsic brightness changes, microlensing 
brightness variations took place during our monitoring time period, this 
might cause certain bias in the time delays estimates, and difference in 
the amplitudes $k_j$ determined from (4). With the assumption that there 
were no strong microlensing events during our monitoring, possible minor 
effects of microlensing at the time interval of 100 days can be 
approximately represented by linear trends (see Fig.\ref{mic_trends}). 
To check a suggestion that the observed amplitude difference results 
from microlensing of the components, we changed Eq. (4), having 
introduced, in addition to the constant terms $a_{0,j}$, linear terms of 
the polynomial as well, $a_{1,j}$, for each component separately:

\begin{equation}
\sum_j\sum_i[k_j f(\vec a, t_i, \tau_j)+a_{0,j}+a_{1,j}(t_i+\tau_j)-
m_j(t_i)]^2=min.
\end{equation}

With this modification of the JFM, which we called JFM$_1$, the averaged 
linear terms of the components would represent a linear term of the 
quasar brightness variation, while their difference might be interpreted 
as the effects of weak microlensings of individual components. If this 
were a case, such a procedure would equalize the amplitudes of the 
components, but this does not happen. 
The amplitudes can also be equalized forcedly by assuming $k_j\equiv1$ 
for all the j-s in condition (6), (JFM$_2$ modification of the JFM). 
After such a procedure, however, residuals in Eq.4 became much 
larger than for the JFM and JFM$_1$,  - the light curves  just stopped 
to be similar, - and this finally rejects a possibility for the 
amplitudes to be equalized by introducing microlensing trends.

Therefore, our assumption that the quasar brightness variation is 
observed in different components with different amplitudes because 
of microlensings does not get a solid confirmation. It should 
be note also, that the time delays calculated with the JFM$_1$ are 
less reliable as compared to those obtained with the JFM, since 
the JFM$_1$ deals with a larger number of unknown parameters. Taking 
all this into account, we placed the results of applying the JFM$_1$ 
and JFM$_2$  in Table~\ref{time_delays} (columns 4 and 5), but did 
not use them in calculations of the average time delays (the bottom 
of Table~\ref{time_delays}). Possible reasons for the observed 
difference in light curves amplitudes will be discussed in Section 5. 
 
A 4-th order polynomials were used for fitting in all cases. Further 
increasing of the polynomial order does not improve the accuracy of 
fitting, and further simulation shows that the random error of time 
delay determination increases slightly as the polynomial degree grows. 

\begin{table}
\caption{Estimates of time delays for Q2237+0305 system with 
respect to the A image, (hours), taken with different approaches from
the Maidanak data, and from those of the OGLE group. The estimates 
averaged over NCCF and JFM for both Maidanak and OGLE data are also
shown, (weighted average values). The JFM$_1$ and JFM$_2$ columns are 
shown just to illustrate the attempt to equalize the amplitudes of 
the light curves, and they were not used to calculate the average 
time delays.} 
\label{time_delays}
\begin{center}
\begin{tabular}[h]{ccccc}
{\it Maidanak data}&&&&\\
\hline
Im. pairs & NCCF & JFM & JMF$_1$  & JFM$_2$\\
\hline
BA     & $-13.2\pm59$  &$-16.1\pm62$  &  33.1 &  -36.2 \\
CA     & $-31.9\pm42$  &$-37.2\pm42$  &  36.9 &  -10.6 \\
DA     &  $6.0\pm60$   & $5.0\pm63$   &  9.1  &  -40.1 \\
\hline
&&&\\
{\it OGLE data}&&&\\
\hline 
BA     &    $1.6\pm52$ &  $5.0\pm57$  &  -  &     7.2 \\
CA     &  $105.8\pm40$ & $105.1\pm43$  &  -  &   117.8 \\
DA     &  $-5.9\pm152$ & $-3.6\pm156$  &  -  &    17.8 \\
\hline
&&&\\
{\it Average}&&&\\
\cline{1-2}
BA  & $-6\pm41$ &   &  &\\
CA  & $35\pm30$ &   &  &\\
DA  & $ 2\pm44$ &   &  &\\
\cline{1-2}
\end{tabular}
\end{center}
\end{table}

\begin{table}
\caption{Model predictions for the time delays $\tau_{BA}$, $\tau_{CA},$ 
and $\tau_{DA}$ for Q2237+0305 system, (hours).}
\label{mod_pred}
\begin{center}
\begin{tabular}[h]{ccccc}
\hline
 \noalign{\smallskip}
 Reference   & Lens model    &$\tau_{BA}$&$\tau_{CA}$&$\tau_{DA}$ \\
 \hline
 \hline
 \raisebox{-1ex}{Schneider,}  &\raisebox{-1ex}{Constant mass-to-light}&  
         &          & \\     
 et al.(1988)&ratio&\raisebox{1.5ex}{-2.4}&\raisebox{1.5ex}{-29.5} 
 &\raisebox{1.5ex}{-26.6} \\ 
 \hline

\raisebox{-1.5ex}{Rix et al.}&Model 1 & 2.1 & 11 & 7.1 \\
\raisebox{-1.5ex}{(1992)}&Model 2 & 1.5 & 10.1 & 6.1 \\
                         &Model 2a &-1.7& 9.8 & 3.7\\
 \hline
 Wambsganss  &Point lens&-2.97    &17.41        &4.87\\
 \& Paczynski &Isothermal sphere &   -1.51  & 8.91 & 2.46 \\
 (1994)  &Best fit &-0.44 & 2.54 & 0.7\\
  \hline
  Schmidt,    &\raisebox{-1ex}{Bar accounted}&\raisebox{-1ex}{-2.0} 
 &\raisebox{-1ex}{16.2} &\raisebox{-1ex} {4.9} \\
 et al.(1996)&           &          &          &     \\
 \hline
\end{tabular}
\end{center}
\end{table}

To determine reliability of estimates, and to analyse the effects 
of random photometric errors, numerical simulations were undertaken. 
The eighth-order polynomial fit to the image A light curve was 
adopted as an intrinsic quasar brightness curve template. Thus we 
admitted, that the actual brightness variation of the quasar might 
be more complicated as compared to the 4-th order polynomial fit 
used for time delay estimations. Individual light curves of the 
components were then constructed by adding to the template curve, 
at the time moments corresponding to those of observations, the 
random noise quantities with the RMS values of $0.015^m$, $0.023^m$, 
$0.016^m$, and $0.021^m$ inherent in photometry for images  A, B, 
C and D, respectively. A zero time lag was adopted for all the 
images, and 2000 random samples for each quaternary of images were 
formed for the simulation. The algorithms described in the previous 
section were applied to the random samples to calculate the average 
time delays with respect to image A, and their RMS deviations. 
The estimates of time delays from these simulated samples did 
not reveal any noticeable biases, and the RMS errors turned out 
to be about 62, 42 and 63 hours for B, C and D, respectively.

We have also considered the possibility that a small amount of 
systematic microlensing could influence these time delay 
determinations. We have simulated the effects of such small 
systematic microlensing as a sustained brightness increase 
(decrease) during 100 days of our monitoring. To do this, the 
actual light curves were distorted by linear trends, and the 
resulting time delay biases were estimated. For our low amplitude 
trends, the biases of the time delay estimates were found to 
depend almost linearly on the value of trend, with 1 day for the 
daily rate of $0.0005^m$. In Sec. 2, (see also Fig. 2) the maximal 
possible value of a microlensing trend has been estimated to be 
about $0.00025^m$ per day. Therefore, the maximum possible time 
delay bias caused by uncompensated microlensing events does 
not exceed 12 hours.

We also tried to calculate the time delays using the OGLE data, 
(http://www.astrouw.edu.pl~ogle/ogle3/huchra.html). The results 
of applying the NCCF and JF methods are shown in Table~\ref
{time_delays}, (columns 2 and 3), as well as the attempts to
equalize the amplitudes of the OGLE light curves, (JFM{$_2$} 
column). The results show, that the high accuracy inherent 
the OGLE photometry, does not seem to improve the results. 
The error estimates resulting from our simulations as above, 
are still very large, and the time delays differ from our 
values and also from the values predicted from available 
macro-imaging models. 

The results of averaging the estimates taken with the NCCF and 
JFM for both the Maidanak and OGLE light curves are also shown 
in Table~\ref{time_delays}, where the weighted average values 
are given. In the following sections, we analyse these results 
and discuss the pitfalls and prospects of observational 
measurement of time delays for the Q2237+0305 system, from daily 
sampled optical wavelength data. 

\section{Time delay predictions for Q2237 from macro-imaging models}

The first predictions of time delays for the Q2237+0305 system appeared 
soon after its identification as a quasar quadruply imaged by the 
gravitational field of a foreground galaxy, (Schneider et al. 
\cite{schn88}, Kent \& Falco \cite{kent88}). Schneider  et al. 
(\cite{schn88}) treated the lensing galaxy projected mass distribution 
as proportional to the observed light distribution and found that 
"a constant mass-to-light ratio, elliptical, de Vaucoulers bulge 
reproduces the observed configuration remarkably well." The model 
was improved in 1992 with the use of a more accurate galaxy light 
distribution using the Hubble Space telescope data (Rix et al. 
\cite{rix92}). Kent and Falco (\cite{kent88}) used another approach 
and considered elliptical mass distributions with a de Vaucouleurs 
and a King profile mass distribution. Another way to account for the 
mass distribution ellipticity was proposed, e.g., by 
Kochanek (\cite{koch91}) and Wambsganss \& Paczynski (\cite{w-p94}). 
They used a shear, $\gamma$, while the mass distribution in the lens 
was adopted to be spherically symmetric, - a singular isothermal 
sphere and a point mass plus different types of the shear, (Kochanek 
\cite{koch91}), and a singular power-law sphere with an external 
shear (Wambsganss \& Paczynski \cite{w-p94}). A massive disk with a 
central nucleus as a point mass, isothermal sphere and Plammer's 
sphere were also considered as models for the Q2237 lensing galaxy, 
(Minakov \& Shalyapin \cite{min91}). An attempt to account for the 
galaxy bar should be mentioned also, (Schmidt et al. \cite{swl98}). 
They used two-dimensional Ferrers profiles for the mass distribution 
in the bar, and a power-law elliptical mass distribution for the bulge, 
with the bar representing a small perturbation of the deflecting field 
of the bulge and constituting about 5\% of the bulge mass in the 
region inside the quasar images.

 Predictions of the time delays for the Q2237 images that we could find 
 in the works cited above, are collected into Table~\ref{mod_pred}, 
with the proper referencing and brief comments concerning each particular
model used. Here, Models 1 and 2 by Rix et al.(\cite{rix92}) mean, 
respectively, $R^{1/4}$ profile, $R^{1/4}$ plus unresolved nucleus, and 
Model 2a means Model 2 with only image positions matched. The predicted 
values of time delays are seen to be obviously model dependent. The 
values of the expected time delays calculated by Wambsganss \& Paczynski 
(\cite{w-p94}) exhibit a tendency to decrease as the index of power 
$\beta$ in their power-law mass profile grows - from $\beta=0$ for a 
"point lens", through $\beta=1$ for an isothermal sphere, and up to 
$\beta=1.71$, ("best fit model").  The B component is always leading, 
in contrast to the model by Rix et al. (\cite{rix92}). A rather good 
agreement of their point lens model predictions with those by Schmidt 
et al. (\cite{swl98}) should be noted, where the effect of the galaxy 
bar was taken into account.

In light of the errors and simulations presented in section 3, we can 
make no conclusions about the comparison of models with our lag 
calculations from monitoring data.

\section {Quasar Intrinsic Brightness Fluctuation Amplitude }

In analysing our calculations, one more thing should be noted and 
discussed. It has been shown in Sec. 2 (see Table~\ref{corr_ind}), 
that the light curves of all the components are highly correlated, 
both in the $R$ filter (our observations), and in the $V$ band 
(OGLE data). It should be noted however that the $R$ light curves  
built from the same Maidanak data by the Moscow group, (Koptelova 
et al. \cite{kop05}) differ from our light curves, with the cross 
correlation indices being much lower than those shown in 
Table~\ref{corr_ind}, and ranging from 0.2 to 0.7. 

\begin{figure}
\resizebox{\hsize}{7cm}{\includegraphics{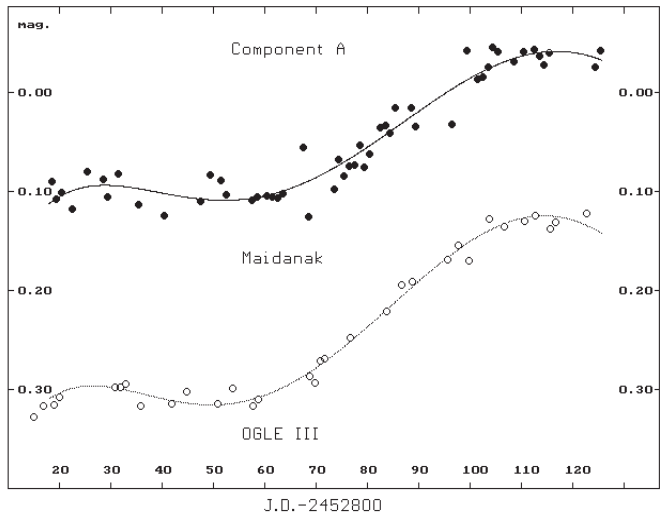}}
 \caption{$R$ light curve of image A from observations on Maidanak in
 June 27 - October 12, 2003, (our photometry), - upper curve, solid 
 circles, - and $V$ light curve  of image A built from the OGLE data, 
 - bottom curve.}  
  \label{OGLE_OUR}
\end{figure}
\begin{table}
\caption{Relative amplitudes of $R$ light curves, (our data), 
 and light curves in $V$ (OGLE data) calculated for the time period 
 June 27  - October  12, 2003.}
\label{ampl}
\begin{center}
\begin{tabular}[h]{ccc}
\hline
 \noalign{\smallskip}
Image &    $R$  &    $V$ \\      
\hline
A  &   1.000    &  1.263 \\        
B  &   0.809    &  0.796 \\    
C  &   1.004    &  0.854 \\          
D  &   0.751    &  0.370 \\ 
\hline         
 \end{tabular}
\end{center}
\end{table} 
If the light curves observed during the time period from June 27 
to October 12, 2003, represent a quasar intrinsic brightness change,  
then its amplitude seen in different lensed images is believed to be 
the same. As seen from Table~\ref{ampl}, where the amplitudes of our 
$R$ light curves and those of the OGLE group in the $V$ band are 
presented in the units of "our" A component amplitude, this is not 
the case. The amplitudes of the A and C light curves are noticeably 
larger than that of D both from our data and from the OGLE group 
data. Inconsistencies between our measurements and the OGLE data 
can also be seen from this table. To better demonstrate what kind 
of inconsistencies between the Maidanak and OGLE data take place, 
- and, at the same time, how similarly they look, - we plotted our 
$R$ and the OGLE $V$ light curves of image A in the same picture, 
(Fig.\ref{OGLE_OUR}).

A possible explanation of these complex differences in fluctuation 
amplitude may be the following. We adopt the picture that all four 
image components may be microlensed but far enough from a  caustic 
in the microlensing pattern that almost no change in relative 
brightnesses of the 4 images is seen during our monitoring period. 
It is also possible that the microlensing caustic pattern in the 
four images is moving approximately parallel to the local caustic 
structure. It is known that microlensing can cause an image to 
become bluer as the caustics cross the quasar structure and magnify 
most strongly the bluest innermost region (Wambsganss \& Paczynski, 
\cite{wp91}; Vakulik et al., \cite{vak04}). Presumably at the 2003 
time of these 
observations, image A was brightest (and image B was faintest) 
because of microlensing image magnification (demagnification). 
Such magnification can alter the color of the images as noted 
previously, and depending upon the details of which part of the 
image are most strongly microlensed, and which part of the quasar 
became brighter in the course of the quasar intrinsic brightness 
fluctuation, color microlensing effects of several kinds might 
be expected. These topics have not been explored with extensive 
simulations to date, because there is as yet no standard quasar 
model. We also note the possibility that the unequal amplitudes 
might result in part from complex structure in either the quasar 
image (Schild \& Vakulik \cite{sv03}),  or in the lens galaxy. 
In the latter case, a subtle unresolved lens galaxy brightness 
feature might cause the erroneous  estimation of an individual 
image brightness; however we would expect this to have been 
noticed in other Q2237 data analysis.

We also note a possible artifact from systematic errors in image 
processing have been  discussed already in a number of 
publications, e.g., Alcalde  et al. (\cite{alc02}), Burud et al. 
(\cite{bur98}); Vakulik et al. (\cite{vak04}). Because of closeness
of quasar images to the galaxy nucleus and to each other, errors 
of processing due to incorrect subtraction of foreground galaxy 
light or due to incorrect subtraction of light from the other 
images are possible, different for different photometry 
methods.

\section{Strategy to Detect Time Delays}

The interesting X-ray observation of possible 2.7 hour time 
delay between the A and B images (Dai et al. 2003) raises the 
question of limits of our ability to measure with precision 
the lags for all four images from optical data. In spite of 
years of Q2237 brightness monitoring, little is yet known 
about the quasar's intrinsic brightness fluctuations on long 
or short time scales.

On short time scales, Burud et al. (\cite{bur98}) and Cumming et al. 
(\cite{cum95}) monitored Q2237 intensively for three nights with the 
NOT and with the CFHT. And nightly monitoring by Vakulik et al. 
(\cite{vak04}) for 46 nights in 2000 with the R filter precludes 
fluctuations of more than $1\%$. 

We can also recall what is known about rapid fluctuations in other 
quasars. Colley \& Schild (\cite{c-s03}) present convincing 
evidence for Q0957 nightly brightness fluctuations at the level of 0.01 
magnitude (their Fig. 1). Gopal-Krishna et al. (\cite{sta03}) report 
detection of intranight variations of brightness of about 0.01 - 0.02 
magnitudes in their sample of radio-quiet quasars. At the same time, 
in a survey of 23 high-luminosity ($-27<M_V<-30$) radio-quiet quasars 
by Rabbette et al. (\cite{rab98}), "no evidence for short-term 
variability greater than about 0.1 magnitudes was detected in any 
of the 23 sources".

Long-term variability of quasars is 
rather well documented with the use of huge amounts of observational 
data, - e.g., results of statistical processing of a sample 
containing over 40000 quasars as reported  by de Vries et al. 
(\cite{Vri05}). Structure function analysis of this sample 
produced estimates of short-term variability consistent with the above,
and also produced 
some important statistical characteristics of the variability 
of quasars, with the following ones relevant here:
1)the increase of variability with increasing time lags is monotonic 
and constant up to time-scales of $\sim 40$ years;
2)variability increases toward the blue part of the spectrum, and
3)high-luminosity quasars vary less than low-luminosity quasars. 

Using this information, and based upon the results of this paper, we 
propose the following strategy for future attempts to observationally 
determine the time delays for the Q2237+0305 gravitationally lensed 
quasar: 
\begin{itemize}
\item since variability  of quasars tends to increase towards the 
shorter wavelength, that is
found for a sample of quasars by de Vries et al. (\cite{Vri05}), 
and also detected for the Q2237 system, (Vakulik et al. \cite{vak04}), 
observations should be shifted towards as much shorter wavelengths 
as possible;
\item monitoring in sets with durations of several weeks and sampling 
interval of hours;
\item observations should be made under the image quality of 
0.6-0.7\arcsec and better, and with the photometry accuracy 
of $0.005^m$ and higher. 
\end{itemize}

\section{Conclusions}

1. Variations of brightness of the four quasar images of 
Q2237, observed during almost 100 days in June - October 
2003, are due mainly to the quasar intrinsic brightness change. 
Microlensing activity was substantially subdued during this 
time period and did not exceed 0.00025 magnitudes per day. 

\noindent 2. The estimates of time delays obtained with the two 
methods proposed, are consistent with each other, but differ 
from those predicted by the existing macro-imaging models. 
The resulting error bars neither confirm nor rule out any of 
the competing models.

\noindent 3. The ultimate limits of time delays estimation from the 
available observations were tested in simulations and have 
shown that, with the particular type of the quasar variability, 
particular photometric errors and contribution from microlensing 
activity, one should not expect better results and higher accuracy. 

\noindent 4. The measured time delays between the 4 quasar images are 
much shorter than the time scales of observed optical microlensing 
brightness changes. Therefore the long history of analysing
brightness changes in the Q2237 literature is based upon a correct 
belief that the observed fluctuations are due mainly to microlensing
events. However, the detected brightness change of the quasar, 
- $0.15^m$ between the Julian dates 860 and 910, i.e. about $0.003^m$ 
per day, - demonstrates that the quasar intrinsic variability may 
noticeably contribute to the observed light curves of the components, 
and thus it should be taken into account when analysing and interpreting 
statistics of microlensing in the system.

\begin{acknowledgements}

The work has been substantially supported by the STCU grants NN43
and U127, and by the joint Ukrainian-Uzbek Program "Development 
of observational base for optical astronomy on Maidanak Mountain". 
We also gratefully acknowledge the use of data obtained by the 
German-Uzbek collaboration between Potsdam/Heidelberg University 
(Robert Schmidt, Joachim Wambsganss), and Astrophysical Institute 
Potsdam (Stefan Gottl\"ober, Lutz Wisotzki), which is supported by 
the Deutsche Forschungsgemeinschaft under grant  436 USB 113/5/0-1.

\end{acknowledgements}

\end{document}